\documentstyle[12pt, a4]{article}
\begin{document}
\author{M. Kocak, I. Zorba, B. G\"{o}n\"{u}l
\and Department of Engineering Physics, Faculty of
Engineering,\and University of Gaziantep, 27310 Gaziantep,
T\"{u}rkiye}
\title{Mapping of non-central potentials under point canonical transformations}
\maketitle
\begin{abstract}
Motivated by the observation that all known exactly solvable shape
invariant central potentials are inter-related via point canonical
transformations, we develop an algebraic framework to show that a
similar mapping procedure is also exist between a class of
non-central potentials. As an illustrative example, we discuss the
inter-relation between the generalized Coulomb and oscillator
systems.
\end{abstract}
Pacs Numbers: 03.65.Fd
\begin{description}
\item[Keywords:] Shape invariance, non-central potential, point canonical
transformation.
\end{description}
\section{Introduction}
In recent years, there has been considerable interest in studying
exactly solvable quantum mechanical problems using algebraic
approaches \cite{chaturvedi}. In this respect, supersymmetric
quantum mechanics (SUSYQM) \cite{cooper} has been found to be an
elegant and useful prescription for obtaining closed analytic
expressions both for the energy eigenvalues and eigenfunctions for
a large class of one-dimensional (or spherically symmetric
three-dimensional) problems. An interesting feature of SUSYQM is
that for a shape invariant system \cite{cooper, infeld} the entire
spectrum can be determined algebraically without ever referring to
underlying differential equations.

As has been shown recently \cite{khare, gonul}, the idea of
supersymmetry and shape invariance can also be used to obtain
exact solutions of a wide class of non-central but separable
potentials in algebraic fashion. In these works, it emerges that
the angular part, as well as the radial part, of the Laplacian of
the Schr\"{o}dinger  equation can indeed be dealt with using the
idea of shape invariance, hence the radial and the angular pieces
of the Schr\"{o}dinger equation can both be treated within the
same framework.

It is well known that the Natanzon potentials \cite{natanzon} are
exactly solvable in non-relativistic quantum mechanics. These
potentials are known to group into two disjoint classes depending
on whether the Schr\"{o}dinger equation can be reduced to either a
hypergeometric or a confluent hypergeometric equation. It has been
shown that \cite{cooper II, dutt, levai} the members within each
class can be mapped into each other by point canonical
transformations (PCT); however members of these two different
classes cannot be connected by PCT. Nevertheless, it is reasonable
to expect that the potentials of the above mentioned two classes
can also be connected by a similar procedure since a
hypergeometric differential equation reduces to a confluent
hypergeometric one under appropriate limits. Gangopadhyaya and his
co-workers have shown \cite{gangopadhyaya} that this is indeed the
case by establishing a connection between the two classes with
appropriate limiting procedures and redefinition of parameters,
thereby inter-relating all known solvable central potentials.

At this stage, it is quite natural to ask whether it is also
possible to inter-connect non-central potentials among themselves
via canonical transformations of coordinates. To the best of our
knowledge, the answer of this question or the feasibility of
application of PCT to non-central potentials for mapping purposes
has not been discussed earlier in the literature. In this respect
such an attempt will be interesting. Through the present article
we will show that the problem posed is algebraically solvable and
mappings are possible between non-central but separable potentials
so long as the separated problems for each of the coordinates
belong to the class of shape invariant potentials (SIP).

The whole development is very elegant, appealing, and yet rather
simple, so that any student of quantum mechanics should be able to
understand and appreciate it. Indeed, we strongly feel that the
material presented here can be profitably included in future
quantum mechanics courses and textbooks. Accordingly, we have kept
this article at a pedagogical level and made it as self-contained
as possible. In the following section, we review briefly PCT in
non-relativistic quantum mechanics. Section 3 explains how the
results for the known SIP may be used to inter-relate two
super-integrable systems: the generalized Coulomb and oscillator
systems. Some concluding remarks are given in the last section.
Throughout the present work the natural units $\hbar=2m=1$  are
used.

\section{Operator transformation}
We consider a time-independent Schr\"{o}dinger equation with a
shape invariant potential $V(\alpha_{i};x)$ that may depend upon
several parameters $\alpha_{i}$
\begin{equation}
\left[-\frac{d^{2}}{dx^{2}}+V(\alpha_{i};x)-E(\alpha_{i})\right]\psi(\alpha_{i};x)=0.
\end{equation}
Under a point canonical transformation, which replaces the
independent variable $x$   by $z(x=f(z))$  and transforms the
wavefunction
$[\psi(\alpha_{i};x)=\nu(z)\tilde{\psi}(\alpha_{i};x)]$ , the
Schr\"{o}dinger equation transforms into:
\begin{eqnarray}
\left[-\frac{d^{2}}{dz^{2}}+\tilde{V}(\tilde{\alpha_{i}};z)-E(\tilde{\alpha_{i}})\right]\tilde{\psi}(\tilde{\alpha_{i}};z)=
\nonumber
\end{eqnarray}
\begin{eqnarray}
-\frac{d^{2}\tilde{\psi}}{dz^{2}}-\left\{\frac{2\nu'}{\nu}-\frac{f''}{f'}\right\}\frac{d\tilde{\psi}}{dz}+\left\{{f'}^{2}\left[V(\alpha_{i};f(z))-E(\alpha_{i})\right]+\left(\frac{f''\nu'}{f'\nu}-\frac{\nu''}{\nu}\right)\right\}\tilde{\psi}=0,
\end{eqnarray}
in which $\tilde{\alpha_{i}}$  represents sets of parameters of
the transformed potentials, and the prime denotes differantion
with respect to the variable $z$ . To remove the first derivative
term in (2) for the purpose of having a Schr\"{o}dinger like
equation, one requires $\nu(z)=c\sqrt{{f'(z)}}$ where $c$ is a
constant of integration. This then leads to another
Schr\"{o}dinger equation with a new potential,
\begin{equation}
\left[-\frac{d^{2}}{dz^{2}}+U(\alpha_{i};z)\right]\tilde{\psi}(\tilde{\alpha_{i}};z)=0,
\end{equation}
where
\begin{equation}
U(\alpha_{i};z)={f'}^{2}[V(\alpha_{i};f(z))-E(\alpha_{i})]+\frac{1}{2}\left(\frac{3f''^{2}}{2f'^{2}}-\frac{f'''}{f'}\right)=\tilde{V}({\tilde{\alpha_{i}};z})-\tilde{E}(\tilde{\alpha_{i}}).
\end{equation}
In general, this is not an eigenvalue equation, unless
$\left\{f'^{2}[V(\alpha_{i};f(z))-E(\alpha_{i})]\right\}$ has a
term independent of $z$, which will act like the energy term for
the new Hamiltonian. This condition constraints allowable choices
for the function $f(z)$. For a general potential $V(z;f(z))$, many
choices for $f(z)$ are still possible that would give rise to
Schr\"{o}dinger type eigenvalue equations, and thus, if we have
one solvable model, we can generate many others from it.

More precisely, the transformation function $f(z)$ has to be
chosen such that the functional form of $U(\alpha_{i};z)$  as
given by (4) is identical to that of the well known exactly
solvable SIP. This is indeed the case if
\begin{equation}
{f'}^{2}[V(\alpha_{i};f(z))-E(\alpha_{i})]=V_{SIP}(\alpha_{i};z)-\frac{1}{2}\left(\frac{3f''^{2}}{2f'^{2}}-\frac{f'''}{f'}\right),
\end{equation}
where $V_{SIP}(\alpha_{i};z)$is a member of the shape invariant
potential family. Ref. \cite{dutt} contains a list of functions
$f(z)$, hence one can easily find the sequence of transformations
necessary to map any shape invariant potential of a given class to
another one belonging to the same class.

We note that PCT have been studied in the path integral approach
to quantum mechanical problems \cite{pak}. However, in path
integral calculations, the mathematical maneuvering of steps
becomes so complicated due to the combined transformations of
space and time variables that the mapping of all SIP has not yet
been done. In this respect, the operator transformation introduced
by Ref. \cite{dutt}, and reviewed above, is a much simpler
approach consists of mapping through canonical transformations of
coordinates which inter-relate the Hilbert spaces of various SIP.
\section{Inter-relations of solvable non-central potentials}
Non-central potentials are normally not discussed in most text
books on quantum mechanics. This is presumably because most of
them are not analytically solvable. However, it is worth noting
that there is a class of non-central potentials in
three-dimensions for which the Schr\"{o}dinger equation separable.
This section deals with the link between two systems, so-called
super-integrable systems, involving a generalized form of such
potentials: a system known in quantum chemistry as the Hartmann
system and a system of potential use in quantum chemistry and
nuclear physics. Both systems correspond to ring-shaped
potentials. They admit two maximally super integrable systems as
the limiting cases: the Coulomb-Kepler system and the isotropic
harmonic oscillator system in three-dimensions. Three-dimensional
potentials that are singular along curves have received a great
deal of attention in recent years. In particular, the Coulombic
ring-shaped potential \cite{kibler} reviewed in quantum chemistry
by Hartmann and co-workers \cite{hartmann}, and the oscillatory
ring-shaped potential \cite{carpido}, systematically studied by
Quesne \cite{quesne}, have been investigated from a quantum
mechanical viewpoint by using various approaches. As ring-shaped
systems they may play an important role in all situations where
axial symmetry is relevant. For example, the Coulomb-Kepler system
is of interest for ring-shaped molecules like cyclic polyenes
\cite{hartmann}. Further, the harmonic oscillator system is of
potential use in the study of (super-) deformed nuclei.
\subsection{The generalized Coulomb system}
For the sake of clarity, and to demonstrate the simplicity of the
present approach, we first review the algebraic framework to solve
exactly the generalized Coulomb system studied in our recent work
\cite{gonul}. This will make clear that how the results for the
known SIP may be used  to algebraically obtain in a closed form
the eigenvalues for a non-central but separable potential. In
addition, as the initial potential to be mapped is the generalized
Coulomb system here, this review would also be very useful in
understanding further the mapping of the generalized Coulomb
system to the oscillatory system having a non-central potential
discussed in the next section.

The Coulombic ring-shaped, or Hartmann, potential (energy) is
\begin{equation}
V=-Z\frac{1}{\sqrt{{x_{1}^{2}}+x_{2}^{2}+x_{3}^{2}}}+\frac{1}{2}Q\frac{1}{{x_{1}^{2}}+x_{2}^{2}}~~~~Z>0~~~~
Q>0,
\end{equation}
where $Z=\eta\sigma^{2}$ and $Q=q\eta^{2}\sigma^{2}$ in the
notation of Kibler and Negadi \cite{kibler} and of Hartmann
\cite{hartmann}. Such an $O(2)$ invariant potential reduces to an
attractive Coulomb potential in the limiting case $Q=0$ and this
will prove useful for checking purposes. Clearly, Eq. (6) is a
special case of the potential (in spherical coordinates)
\begin{equation}
V_{GC}(r,\theta)=\frac{A}{r}+\frac{B}{r^{2}\sin^{2}\theta}+C\frac{\cos\theta}{r^{2}\sin^{2}\theta},
\end{equation}
introduced by Makarov et al. \cite{makarov}. The importance of the
potential in (7) lies on the fact that compound Coulomb plus
Aharanov-Bohm potential \cite{aharonov} and Hartmann ring-shaped
potential, originally proposed as model for the benzene molecule
are mathematically linked to this potential. In fact the energy
spectrum for these two potentials can be obtained directly
\cite{gonul} by considering these as a special case of the general
non-central potential in (7) .

The Schr\"{o}dinger equation in spherical polar coordinates for a
particle in the presence of a potential $V_{GC}(r,\theta)$ can be
reduced to the two ordinary differential equations,
\begin{equation}
\frac{d^{2}R}{dr^{2}}+\frac{2}{r}\frac{dR}{dr}+\left(E-\frac{A}{r}-\frac{l(l+1)}{r^{2}}\right)R=0,
\end{equation}
\begin{equation}
\frac{d^{2}P}{d\theta^{2}}+\cot\theta\frac{dP}{d\theta}+\left[l(l+1)-\frac{m^{2}}{\sin^{2}\theta}-\frac{(B+C\cos\theta)}{\sin^{2}\theta}\right]P=0,
\end{equation}
if the corresponding total wave function can be written as
$\psi(r,\theta,\vartheta)=R(r)P(\theta)\exp^{im\vartheta}$. It is
not difficult to see that Eq. (8) is the same we obtain in solving
the problem of an electron in a Coulomb-like field. Bearing in
mind the discussion given in the previous section and using the
transformation $\theta\rightarrow z$ through a mapping function
$\theta=f(z)$, one obtains
\begin{equation}
\frac{d^{2}P}{dz^{2}}+\left[-\frac{f''}{f'}+f'\cot{f}\right]\frac{dP}{dz}+f'^{2}\left[l(l+1)-\frac{m^{2}}{\sin^{2}f}-\frac{(B+C\cos{f})}{\sin^{2}f}\right]P=0,
\end{equation}
which seems a Schr\"{o}dinger-like equation if
$\frac{f''}{f'}=f'\cot{f}$ that leads to
\begin{equation}
\theta\equiv~f~=2\tan^{-1}(e^{z}),~~\sin\theta=\frac{1}{\cosh{z}},~~\cos\theta=-\tanh{z}.
\end{equation}
Eq. (9) now reads
\begin{equation}
\frac{d^{2}P}{dz^{2}}+\left[\frac{l(l+1)}{\cosh^{2}{z}}+C\tanh{z}\right]P=(m^{2}+B)P,
\end{equation}
which can be rearranged as
\begin{equation}
-\frac{d^{2}P}{dz^{2}}+\left[\lambda^{2}\tanh^{2}{z}-C\tanh{z}\right]P=\left[\lambda^{2}-(m^{2}+B)\right]P,
\end{equation}
where $\lambda^{2}=l(l+1)$. The full potential in (13) has the
form of the Rosen-Morse-II  potential which is well-known to be
shape invariant. More specifically, the potential
\begin{equation}
V_{RM}(z)=a_{0}(a_{0}+1)\tanh^{2}z+2b_{0}\tanh{z}~~~~~(b_{0}<a^{2}_{0}),
\end{equation}
has energy eigenvalues \cite{cooper}
\begin{equation}
E_{RM}=a_{0}(a_{0}+1)-(a_{0}-n)^{2}-\frac{b^{2}_{0}}{(a_{0}-n)^{2}}~~~~~~n=0,1,2,...
\end{equation}
For our case $E_{RM}=\lambda^{2}-(m^{2}+B)$ and using the
corresponding constants $a_{0}=l$, $b_{0}=-\frac{C}{2}$ we obtain
\begin{equation}
l=n+\left[\frac{(m^{2}+B)+\sqrt{(m^{2}+B)^{2}-C^{2}}}{2}\right]^{\frac{1}{2}}.
\end{equation}
The energy eigenvalues obtained from (8) for the Coulomb potential
($A=-Ze^{2}$) are
\begin{equation}
E_{C}=\frac{-Z^{2}e^{4}}{4[N+l+1]^{2}}~~~~N=0,1,2,...
\end{equation}
Therefore our final eigenvalues for a bound electron in a Coulomb
potential as well a combination of a non-central potentials given
by (9) are
\begin{equation}
E_{GC}=\frac{-Z^{2}e^{4}}{4\left\{N+n+1+\left[\frac{(m^{2}+B)+\sqrt{(m^{2}+B)^{2}-C^{2}}}{2}\right]^{\frac{1}{2}}\right\}^{2}}.
\end{equation}
\subsection{Mappings between the two distinct systems}
Now, we are ready to illustrate the mapping procedure by starting
off with the system described in (7). We show here the steps
necessary to relate the generalized Coulomb system discussed above
to a system having a generalized oscillatory potential, so called
the generalized Aharonov-Bohm plus oscillator systems,
\begin{equation}
V_{GHO}(r,\theta)={\tilde{A}}r^{2}+\frac{{\tilde{B}}}{(r\cos\theta)^{2}}+\frac{\tilde{C}}{(r\sin\theta)^{2}},
\end{equation}
where $\tilde{A},\tilde{B},\tilde{C}$ being strictly positive
constants. The potential above is of the $V_{3}$ in the
$V_{1}-V_{4}$ classification by Makarov and collaborators
\cite{makarov}. The limiting case $\tilde{B}=0$,~~$\tilde{C}=0$
corresponds to an isotropic harmonic oscillator and will serve for
testing results to be obtained. In case when $\tilde{B}=0$ we get
the well known ring-shape oscillator potential which was
investigated in many articles.

The strategy followed is to start the transformation with
appropriate Schr\"{o}dinger equations which must be exactly
solvable having a shape invariant potential, explicitly these are
Eqs. (8) and (13), and to see what happens to these equations
under a point canonical transformation. In order for the
Schr\"{o}dinger equation to be mapped into another Schr\"{o}dinger
equation, there are severe restrictions on the nature of the
coordinate transformation. Coordinate transformations which
satisfy these restrictions give rise to new solvable problems.
When the relationship between coordinates is implicit, then the
new solutions are only implicitly determined, while if the
relationship is explicit then the newly found solvable potentials
are also shape invariant which will be the case in the present
work.

As the first step, we proceed with the transformation of the well
known system in (8) that involves the central portion of the
non-central potential in (7). For convenience, we will call our
initial coordinates $r$ and our final coordinates $z$. Using the
point canonical transformation $r\equiv f(z)=z^{2}$, one readily
obtains from Eq. (4)
\begin{equation}
U_{1}=-4Ze^{2}+\frac{16l(l+1)+3}{4z^{2}}-(4E_{C})z^{2}.
\end{equation}
where $E_{C}$ is the energy eigenvalue in (17) corresponding to
the initial shape invariant Coulomb potential. As the angular
momentum barrier term in (20) can algebraically be expressed in
the form
\begin{equation}
\frac{16l(l+1)+3}{4z^{2}}=\frac{\tilde{l}(\tilde{l}+1)}{z^{2}},~~~~
\tilde{l}=2l+\frac{1}{2},
\end{equation}
the consideration of Eq. (20) together with the right hand side of
Eq. (4) yields
\begin{equation}
U_{1}=\frac{w^{2}z^{2}}{4}+\frac{\tilde{l}(\tilde{l}+1)}{z^{2}}-2w(N+l+1)=\tilde{V_{1}}-\tilde{E_{1}},
\end{equation}
where the relation between the original and transformed potential
parameters is $w=2Ze^{2}/(N+l+1)$. It is obvious that the
transformed new potential, which represents the first term in (19)
where $\tilde{A}=w^{2}/4$, is the shape invariant isotropic
harmonic oscillator potential
\begin{equation}
\tilde{V}_{1}=\tilde{A}z^{2}+\frac{\tilde{l}(\tilde{l}+1)}{z^{2}},
\end{equation}
and the corresponding energy eigenvalues are
\begin{equation}
\tilde{E}_{1}=w\left(2N+\tilde{l}+\frac{3}{2}\right).
\end{equation}

The next step is to transform the system in (13) which is
identical to (9) involving the angular dependent potentials in
(7). This mapping will enable us to see clearly the final form of
the transformed (new) non-central potential, and to derive an
explicit expression (like Eq. (16) for $\tilde{l}$ appeared in
(24) in terms of the transformed potential parameters of the
non-central portion.

Following a similar algebraic treatment as above, and using a
proper transformation function
\begin{equation}
z\equiv~f~(\tilde{z})=\tanh^{-1}(\cosh2{\tilde{z}}),
\end{equation}
where $z$ and $\tilde{z}$ denote the initial and final coordinates
respectively, one can readily transform the radial Schr\"{o}dinger
equation in (13) and obtain from Eqs. (25) and (4)
\begin{equation}
U_{2}=(2l+1)^{2}-\frac{\left(m^{2}+B+C-\frac{1}{4}\right)}{\cosh^{2}\tilde{z}}+\frac{\left(m^{2}+B-C-\frac{1}{4}\right)}{\sinh^{2}\tilde{z}}=\tilde{V}_{2}-\tilde{E}_{2}.
\end{equation}
This encouraging result complies with Eq.(23) of Ref. \cite{gonul}
where generalized oscillator system Schr\"{o}dinger equations have
been solved using the usual procedure that applied to the
generalized Coulomb system in section 3.1.

In the above equation, the full potential
$\tilde{V}_{2}(\tilde{z})$ resembles the shape invariant
P\"{o}schl-Teller-II type potential
\begin{equation}
V_{PT}(\tilde{z})=-\frac{\tilde{a}_{0}(\tilde{a}_{0}+1)}{\cosh^{2}\tilde{z}}+\frac{\tilde{b}_{0}(\tilde{b}_{0}-1)}{\sinh^{2}\tilde{z}},
\end{equation}
with the eigenenergies
\begin{equation}
E_{PT}=-(\tilde{a}_{0}-\tilde{b}_{0}-2n)^{2}.
\end{equation}
Comparing the similar terms in Eqs. (26-28), and bearing in mind
the relation $\tilde{l}=2l+1/2$ from  (21), one finds
\begin{eqnarray}
\tilde{a}_{0}=-\frac{1}{2}+\sqrt{m^{2}+B+C},~~\tilde{b}_{0}=\frac{1}{2}+\sqrt{m^{2}+B-C},
\nonumber
\end{eqnarray}
\begin{eqnarray}
\tilde{l}=2n+\frac{1}{2}+\sqrt{m^{2}+B-C}+\sqrt{m^{2}+B+C}.
\end{eqnarray}
Note that to ensure the natural relation between $l$ and
$\tilde{l}$, positive roots are chosen for $\tilde{a}_{0}$
and $\tilde{b}_{0}$ values. This choice leads to involve only the
odd solutions \cite{gonul, kibler II}, which means that the states
corresponding to the potential in (7) are mapped onto the
odd-integer states of the transformed potential.

Thus, Eq. (24) reads
\begin{equation}
\tilde{E}_{GHO}=w(2N+2n+2+\sqrt{m^{2}+B-C}+\sqrt{m^{2}+B+C}),~~N=1,3,5,...
\end{equation}
which are the eigenvalues of the transformed non-central potential
formed as in (19) where
\begin{equation}
\tilde{B}=m^{2}+B-C-\frac{1}{4},~~\tilde{C}=B+C,
\end{equation}
The final form of the transformed potential in Eq. (19) can easily
be seen by the use of inverse transformation of the potential in
(26) via the mapping functions in Eq.(11). Eq. (30), which agrees
with Eq. (3) of Ref. \cite{kibler II} and Eq. (27) of Ref.
\cite{gonul}. Thus, we have in this section shown that under the
canonical transformations energies and coupling constants of the
non-central potentials of interest trade places and orbital
angular momenta are re-scaled, which clarifies that there is
really only one quantum mechanical problem, not two.

We finally remark that although here we have only focused on
eigenvalues and spherical polar coordinates, generalization of the
technique used to describe eigenfunctions of the present systems
in analytical form (using the discussion in section 2 and the
analytical forms of the unnormalized wave functions in Refs.
\cite{cooper, dutt}, and to other non-central potentials in any
orthogonal curvilinear coordinate system is quite straightforward.
\section{Concluding Remarks}
In this article the Schr\"{o}dinger equation with a class of
non-central but separable potentials has been studied and we have
shown that such potentials can be easily inter-related among
themselves within the framework of point canonical coordinate
transformations, as the corresponding eigenvalues may be written
down in a closed form algebraically using the well known results
for the shape invariant potentials. Although the literature
covered similar problems, to our knowledge an investigation such
as the one we have discussed in this paper was missing. It is
quite obvious that similar mapping procedures can be followed
starting from other types of shape invariant potentials which may
lead to solve analytically other complicated systems involving
different kind of non-central potentials. With the above
considerations the authors hope to stimulate further examples of
applications of the present method in important problems of
physics.

\newpage\


\begin{thebibliography}{99}
\bibitem{chaturvedi} Chaturvedi S, Dutt R, Gangopadhyaya A, Panigrahi P, Rasinariu C , Sukhatme U 1998 {\it Phys. Lett.} {\bf A 248} 109, and the references therein.
\bibitem{cooper} Cooper F, Khare A, Sukhatme U  1995  {\it Phys. Rep.} {\bf 251} 268.
\bibitem{infeld} Infeld L, Hull T E 1951 {\it Rev. Mod. Phys.} {\bf 23} 21; Gendenshtein L E 1983 {\it JETP Lett.} {\bf 38} 356;
                 Gendenshtein L E, Krive I V 1985 {\it Sov. Phys.} Usp. {\bf 28} 645.
\bibitem{khare}  Khare A, Bhaduri R K {\it hep-th/9310104}; Dutt R, Gangopadhyaya A, Sukhatme U P 1997 {\it Am. J. Phys.} {\bf 65} 400.
\bibitem{gonul}  Gonul B, Zorba I  2000 {\it Phys. Lett.} {\bf A 269} 83.
\bibitem{natanzon}  Natanzon G A 1979 {\it Teor. Mat. Fiz.} {\bf 38} 219.
\bibitem{cooper II}  Cooper F, Ginocchio J N, Wipf A 1989 {\it J. Phys. A: Math. Gen.} {\bf 22} 3707.
\bibitem{dutt} De R, Dutt R, Sukhatme U 1992 {\it J. Phys. A: Math. Gen.} {\bf 25} L843.
\bibitem{levai} Levai G  May 17-19 1993 " Solvable Potentials Derived >From Supersymmetric Quantum Mechanics"
                Talk presented at International Symposium on Quantum Inversion Theory and Applications,
                Bad Honnef, Germany.
\bibitem{gangopadhyaya}  Gangopadhyaya A, Panigrahi P K, Sukhatme U P 1994 {\it Helv. Phys. Acta} {\bf 67} 363.
\bibitem{pak}  Pak N K, Sokmen I 1984 {\it Phys. Lett.} {\bf A 103} 298; Khandekar D C,
               Lawande S V 1986 {\it Phys. Rep.} {\bf 137} 115; Pak N K, Sokmen I 1984 {\it Phys. Rev.} {\bf A 30} 1629;
               Inomata A 1982 {\it Phys. Lett.} {\bf A 87} 387.
\bibitem{kibler}  Kibler M, Negadi T 1984 {\it Int. J. Quantum Chem.} {\bf 26} 405; Sökmen I 1986
                  {\it Phys. Lett.} {\bf A 118} 249; Kibler M, Winternitz P  1987  {\it J. Phys.} A: Math. Gen. {\bf 20} 4097; Lutsenko I V et. al. 1990 Teor. Mat. Fiz. {\bf 83} 419.
\bibitem{hartmann}  Hartmann H et. al. 1972 {\it Theor. Chim. Acta} {\bf 24} 201; Hartmann H, Schuch D 1980 {\it Int. J. Quantum Chem.} {\bf 18} 125.
\bibitem{carpido}  Carpido-Bernido M V, Bernido C C 1989 {\it Phys. Lett.} {\bf A} 134 315; Carpido-Bernido M V
                   1991 {\it J. Phys. A: Math. Gen.} {\bf 24} 3013; Gal'bert O F, Granovskii Y I, Zhedanov A S 1991
                   {\it Phys. Lett.}{\bf A 153} 177.
\bibitem{quesne}  Quesne C 1988 {\it J. Phys. A: Math. Gen.} {\bf 21} 3093.
\bibitem{makarov}  Makarov A A et al. 1967 {\it Nuovo Cimento} {\bf A 52} 1061.
\bibitem{aharonov}  Aharonov Y, Bohm D  1959 {\it Phys. Rev.} {\bf 115} 485.
\bibitem{kibler II}  Kibler M, Campiogotta C 1993 {\it Phys. Lett.} {\bf A} 181 1.
\end{thebibliography}
\end{document}